\documentclass{article}

\usepackage{arxiv}

\usepackage[utf8]{inputenc} 
\usepackage[T1]{fontenc}    
\usepackage{hyperref}       
\usepackage{url}            
\usepackage{booktabs}       
\usepackage{amsfonts}       
\usepackage{amssymb}        
\usepackage{amsmath}        
\usepackage{nicefrac}       
\usepackage{microtype}      
\usepackage{lipsum}		
\usepackage{graphicx}
\usepackage[numbers]{natbib}
\usepackage{doi}
\usepackage{xcolor}
\usepackage{wrapfig}

\title{Disordered vectors in R: introducing the {\tt disordR} package}

\author{ \href{https://orcid.org/0000-0001-5982-0415}{\includegraphics[width=0.03\textwidth]{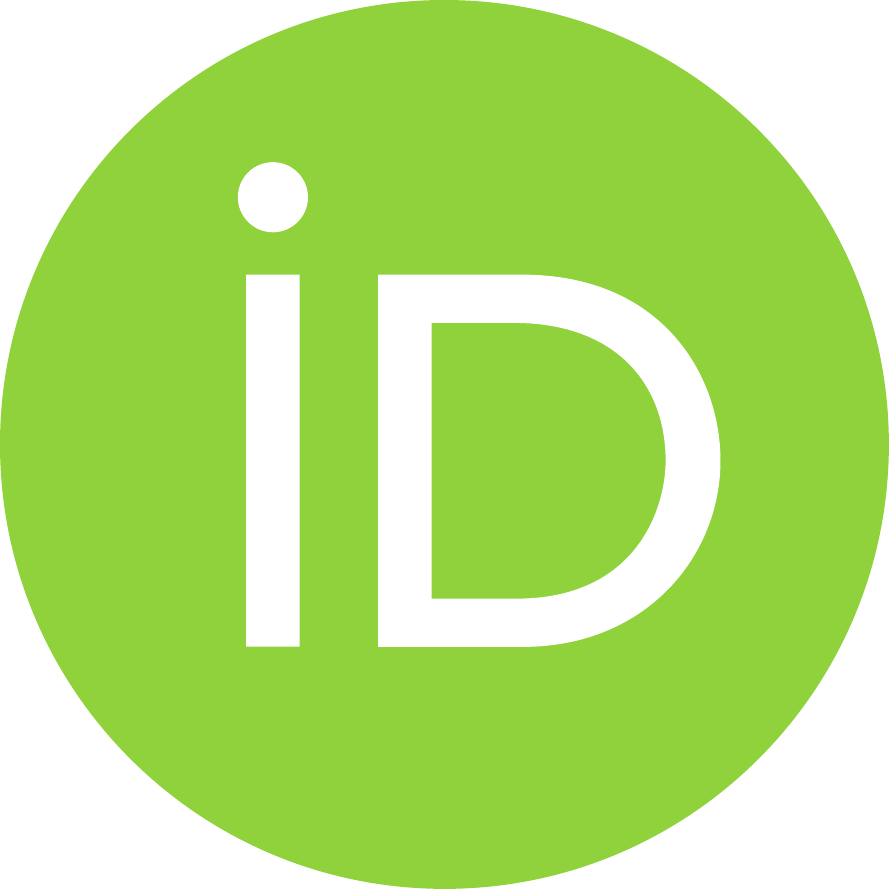}\hspace{1mm}Robin K. S.~Hankin}\thanks{\href{https://academics.aut.ac.nz/robin.hankin}{work};  
\href{https://www.youtube.com/watch?v=JzCX3FqDIOc&list=PL9_n3Tqzq9iWtgD8POJFdnVUCZ_zw6OiB&ab_channel=TrinTragulaGeneralRelativity}{play}} \\
 Auckland University of Technology\\
	\texttt{hankin.robin@gmail.com} \\
}

\renewcommand{\shorttitle}{The \textit{disordR} package}

\hypersetup{
pdftitle={The disordR package},
pdfsubject={q-bio.NC, q-bio.QM},
pdfauthor={Robin K. S.~Hankin},
pdfkeywords={disord vectors}
}

\usepackage{Sweave}
\begin{document}
\maketitle

\begin{abstract}

Objects in the {\tt stl map} class of {\tt C++} associate a value to
each of a set of keys.  Accessing values or keys of such an object is
problematic in the R programming language because the value-key pairs
are not stored in a well-defined order.  This document motivates and
discusses the concept of ``disordered vector" as implemented by the
{\tt disordR} package which facilitates the handling of {\tt map}
objects.  Values and keys of a map are stored in an
implementation-specific way so certain extraction and replacement
operations should be forbidden.  For example, if values are real, then
the ``first" value is implementation specific\ldots but the maximum
value has a well-defined result.  The {\tt disordR} package makes
forbidden operations impossible while allowing transparent R idiom for
permitted operations.  An illustrative R session is given in which the
package is used abstractly, without reference to any particular
application, and then shows how it can be used to manipulate
multivariate polynomials.  The {\tt disordR} package is a dependency
of {\tt clifford}, {\tt freealg}, {\tt hyper2}, {\tt mvp}, {\tt
  spray}, {\tt stokes}, and {\tt weyl}.  The {\tt disordR} package is
available on CRAN at \url{https://CRAN.R-project.org/package=disordR}.
\end{abstract}

\section{Introduction}

\setlength{\intextsep}{0pt}
\begin{wrapfigure}{r}{0.2\textwidth}
  \begin{center}
\includegraphics[width=1in]{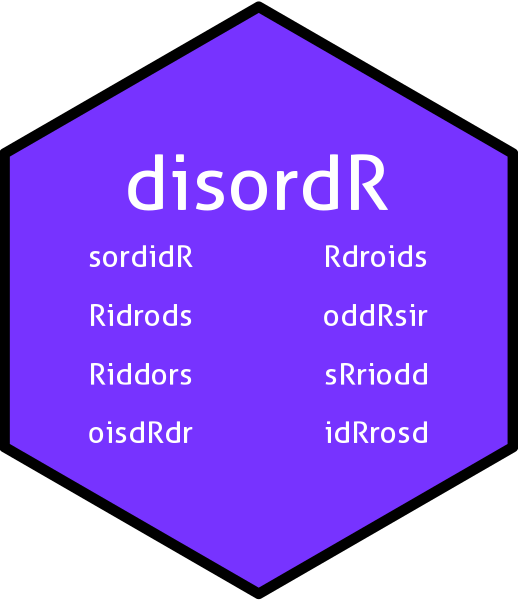}
  \end{center}
\end{wrapfigure}
In {\tt C++}~\cite{ISO14882}, the {\tt stl map}
class~\cite{josuttis1999} is an object that associates a value to each
of a set of keys.  Accessing values or keys of a {\tt map} object is
problematic because the value-key pairs are not stored in a
well-defined order.  The situation is applicable to any package which
uses the {\tt map} objects.  Consider, for example, the {\tt mvp}
package~\citep{hankin2022mvp} which deals with multinomials using
{\tt stl} maps.  An {\tt mvp} object is a map from terms to
coefficients, and a map has no intrinsic ordering: the maps

\verb|x -> 1, xy -> 3, xy^3 -> 4| \hfill and\hfill
\verb|xy^3 -> 4, xy -> 3, x -> 1|

are the same map and correspond to the same object (symbolically,
$x+3xy+4xy^3=4xy^3+3xy+x$).  Thus the coefficients of the multinomial
might be \verb+c(1,3,4)+ or \verb+c(4,3,1)+, or indeed any ordering.
Internally, the elements are stored in some order but the order used
is implementation-specific.  Quite often, I am interested in the
coefficients {\em per se}, without consideration of their meaning in
the context of a multivariate polynomial.  I might ask:

\begin{itemize}
\item  ``How many coefficients are there?"
\item  ``What is the largest coefficient?"
\item  ``Are any coefficients exactly equal to one?"
\item  ``How many coefficients are greater than 2?"
\end{itemize}

These are reasonable and mathematically meaningful questions.  Compare
a meaningless question: ``what is the first coefficient?".  This is
meaningless because of the order ambiguity discussed above: the answer
is at best implementation-specific, but fundamentally it is a question
that one should not be allowed to ask.

To deal with the coefficients in isolation in R~\cite{rcore2022}, one
might be tempted to use a multiset.  However, this approach does not
allow one to link the coefficients with the terms.  Suppose I coerce
the coefficients to a multiset object (as per the {\tt sets}
package~\cite{meyer2009,meyer2022}, for example): then it is
impossible to extract the terms with coefficient greater than 2 (which
would be the polynomial $3xy+4xy^3$) because the link between the
coefficients and the terms is not included in the multiset object.
Sensible questions involving this aspect of {\tt mvp} objects might
be:

\begin{itemize}
\item Give me all terms with coefficients greater than 2
\item Give me all terms with positive coefficients
\item Give me all terms with integer coefficients
\end{itemize}

and these questions cannot be answered if the the coefficients are
stored as a multiset (compare inadmissible questions such as ``give me
the first three terms").  Further note that replacement methods are
mathematically meaningful, for example:

\begin{itemize}
\item Set any term with a negative coefficient to zero
\item Add 100 to any coefficient less than 30
\end{itemize}

Again these operations are reasonable but precluded by multiset
formalism (compare inadmissible replacements: ``replace the first two
terms with zero", or ``double the last term" would be inadmissible).

{\bf\em\Large What we need is a system that forbids stupid questions
and stupid operations, while transparently permitting sensible
questions and operations}

The {\tt disord} class of the {\tt disordR} package is specificially
designed for this situation.  This class of object has a slot for the
coefficients in the form of a numeric R vector, but also another slot
which uses hash codes to prevent users from misusing the ordering of
the numeric vector.
   
For example, a multinomial $x+2y+3z$ might have coefficients {\tt
c(1,2,3)} or {\tt c(3,1,2)}.  Package idiom to extract the
coefficients of a multivariate polynomial {\tt a} is {\tt coeffs(a)};
but this cannot return a standard numeric vector.  If stored as a
numeric vector, the user might ask ``what is the first element?" and
this question should not be asked [and certainly not answered!],
because the elements are stored in an implementation-specific order.
The {\tt disordR} package uses {\tt disord} objects which are designed
to return an error if such inadmissible questions are asked.  But {\tt
disord} objects can answer admissible questions and perform admissible
operations.

Suppose we have two multivariate polynomials, {\tt a} as defined as
above with $a=x+2y+3z$ and $b=x+3y+4z$.  Even though the sum $a+b$ is
well-defined algebraically, idiom such as {\tt coeffs(a) + coeffs(b)}
is not defined because there is no guarantee that the coefficients of
the two multivariate polynomials are stored in the same order.  We
might have {\tt c(1,2,3)+c(1,3,4)=c(2,5,7)} or {\tt
c(1,2,3)+c(1,4,3)=c(2,6,6)}, with neither being more ``correct" than
the other.  In the package, this ambiguity is rendered void: {\tt
coeffs(a) + coeffs(b)} will return an error.  Note carefully that {\tt
coeffs(a+b)} is perfectly well defined, although the result is subject
to the same ambiguity as {\tt coeffs(a)}.

In the same way, {\tt coeffs(a) + 1:3} is not defined and will return
an error.  Further, idiom such as {\tt coeffs(a) <- 1:3} and {\tt
coeffs(a) <- coeffs(b)} are not defined and will also return an error.
However, note that

\begin{Schunk}
\begin{Sinput}
coeffs(a) + coeffs(a)
coeffs(a) + coeffs(a)^2
coeffs(a) <- coeffs(a)^2
coeffs(a) <- coeffs(a)^2 + 7
\end{Sinput}
\end{Schunk}

are perfectly well defined, with package idiom behaving as expected.
In the assignments, one does not need to know the order of the left
hand side, so long as the order is the same on both sides.  The
idiomatic English equivalent would be: ``the coefficient of each term
of {\tt a} becomes its square"; note that this operation is
insensitive to the order of coefficients.  The whole shebang is
intended to make idiom such as {\tt coeffs(a) <- coeffs(a)\%\%2}
possible, so we can manipulate polynomials over finite rings, here
$\mathbb{Z}/2\mathbb{Z}$.

The replacement methods are defined so that an expression like {\tt
coeffs(a)[coeffs(a) < 5] <- 0} works as expected; the English idiom
would be ``replace any coefficient less than 5 with 0".  To fix ideas,
consider a fixed small mvp object:

\begin{Schunk}
\begin{Sinput}
R> library("mvp")
R> a <- as.mvp("5 a c^3 + a^2 d^2 f^2 + 4 a^3 b e^3 + 3 b c f + 2 b^2 e^3")
R> a
\end{Sinput}
\begin{Soutput}
mvp object algebraically equal to
5 a c^3  +  a^2 d^2 f^2  +  4 a^3 b e^3  +  3 b c f  +  2 b^2 e^3
\end{Soutput}
\end{Schunk}

Extraction presents issues; consider {\tt coeffs(a) < 3}.  This object
has Boolean elements but has the same ordering ambiguity as {\tt
coeffs(a)}.  One might expect that we could use this to extract
elements of {\tt coeffs(a)}: specifically, those elements less than 5.
We may use replace methods for coefficients if this makes sense.
Idiom such as

\begin{Schunk}
\begin{Sinput}
R> coeffs(a)[coeffs(a)<5] <- 4 + coeffs(a)[coeffs(a)<5]
R> coeffs(a) <- pmax(coeffs(a),3)
\end{Sinput}
\end{Schunk}

is algebraically meaningful and allowed in the package.
Idiomatically: ``Add 4 to any element less than 5"; ``coefficients
become the parallel maximum of themselves and 3" respectively.
Further note that {\tt coeffs(a) <- rev(coeffs(a))} is disallowed
(although {\tt coeffs(a) <- rev(rev(coeffs(a)))} is meaningful and
admissible).

So the output of {\tt coeffs(x)} is defined only up to an unknown
rearrangement.  The same considerations apply to the output of {\tt
vars()}, which returns a list of character vectors in an undefined
order, and the output of {\tt powers()}, which returns a numeric list
whose elements are in an undefined order.  However, even though the
order of these three objects is undefined individually, their ordering
is jointly consistent in the sense that the first element of {\tt
coeffs(x)} corresponds to the first element of {\tt vars(x)} and the
first element of {\tt powers(x)}.  The identity of this element is not
defined---but whatever it is, the first element of all three accessor
methods refers to it.

Note also that a single term (something like \verb+4a^3*b*c^6+) has
the same issue: the variables are not stored in a well-defined order.
This does not matter because the algebraic value of the term does not
depend on the order in which the variables appear and this term would
be equivalent to \verb+4bc^6*a^3+.

\section{An R session with the {\tt disordR} package}

We will use the {\tt disordR} package to show how the idiom works.

\begin{Schunk}
\begin{Sinput}
R> library("disordR")
R> set.seed(0)
R> a <- rdis()
R> a
\end{Sinput}
\begin{Soutput}
A disord object with hash 5b7279f3c05d00cf1e8f999a755151e0451c56ec and elements
[1] 9 4 7 1 2 6 3 8 5
(in some order)
\end{Soutput}
\end{Schunk}

Object {\tt a} is a {\tt disord} object but it behaves similarly to a
regular numeric vector in many ways:

\begin{Schunk}
\begin{Sinput}
R> a^2
\end{Sinput}
\begin{Soutput}
A disord object with hash 5b7279f3c05d00cf1e8f999a755151e0451c56ec and elements
[1] 81 16 49  1  4 36  9 64 25
(in some order)
\end{Soutput}
\begin{Sinput}
R> a+1/a
\end{Sinput}
\begin{Soutput}
A disord object with hash 5b7279f3c05d00cf1e8f999a755151e0451c56ec and elements
[1] 9.111111 4.250000 7.142857 2.000000 2.500000 6.166667 3.333333
[8] 8.125000 5.200000
(in some order)
\end{Soutput}
\end{Schunk}

Above, note how the result has the same hash code as {\tt a}.  Other
operations that make sense are {\tt max()} and {\tt sort()}:

\begin{Schunk}
\begin{Sinput}
R> max(a)
\end{Sinput}
\begin{Soutput}
[1] 9
\end{Soutput}
\begin{Sinput}
R> sort(a)
\end{Sinput}
\begin{Soutput}
[1] 1 2 3 4 5 6 7 8 9
\end{Soutput}
\end{Schunk}

Above, see how the result is a standard numeric vector.  However,
inadmissible operations give an error.  For example, we cannot extract
the ``first'' element of {\tt a}:

\begin{Schunk}
\begin{Sinput}
R> try(a[1])
\end{Sinput}
\begin{Soutput}
Error in .local(x, i, j = j, ..., drop) : 
  if using a regular index to extract, must extract each element once and once only (or none of them)
\end{Soutput}
\end{Schunk}

nor can we replace it:

\begin{Schunk}
\begin{Sinput}
R> try(a[1] <- 1000)
\end{Sinput}
\begin{Soutput}
Error in .local(x, i, j = j, ..., value) : 
  if using a regular index to replace, must specify each element once and once only
\end{Soutput}
\end{Schunk}

However, the package is designed so that standard R operations
generally work as expected for permissible operations:

\begin{Schunk}
\begin{Sinput}
R> x <- a + 1/a
R> x
\end{Sinput}
\begin{Soutput}
A disord object with hash 5b7279f3c05d00cf1e8f999a755151e0451c56ec and elements
[1] 9.111111 4.250000 7.142857 2.000000 2.500000 6.166667 3.333333
[8] 8.125000 5.200000
(in some order)
\end{Soutput}
\begin{Sinput}
R> y <- a*2-9
R> y
\end{Sinput}
\begin{Soutput}
A disord object with hash 5b7279f3c05d00cf1e8f999a755151e0451c56ec and elements
[1]  9 -1  5 -7 -5  3 -3  7  1
(in some order)
\end{Soutput}
\begin{Sinput}
R> x+y
\end{Sinput}
\begin{Soutput}
A disord object with hash 5b7279f3c05d00cf1e8f999a755151e0451c56ec and elements
[1] 18.1111111  3.2500000 12.1428571 -5.0000000 -2.5000000  9.1666667
[7]  0.3333333 15.1250000  6.2000000
(in some order)
\end{Soutput}
\end{Schunk}

Above, observe that objects {\tt a}, {\tt x} and {\tt y} have the
same hash code: they are ``compatible", in {\tt disordR} idiom.
However, if we try to combine object {\tt a} with another object with
different hash, we get errors:

\begin{Schunk}
\begin{Sinput}
R> (b <- rdis())
\end{Sinput}
\begin{Soutput}
A disord object with hash 488e1c6f4e2c062379d47b5511730a9785661318 and elements
[1] 2 3 8 1 5 6 9 7 4
(in some order)
\end{Soutput}
\begin{Sinput}
R> try(a+b)
\end{Sinput}
\begin{Soutput}
a + b
Error in check_matching_hash(e1, e2, match.call()) : 
hash codes 5b7279f3c05d... and 488e1c6f4e2c... do not match
\end{Soutput}
\end{Schunk}

The error is given because objects {\tt a} and {\tt b} are stored in
an implementation-specific order (we say that {\tt a} and {\tt b} are
{\em incompatible}).  In the package, many extract and replace methods
are implemented whenever this is admissible:

\begin{Schunk}
\begin{Sinput}
R> a[a<0.5] <- 0  # round down
R> a
\end{Sinput}
\begin{Soutput}
A disord object with hash 5b7279f3c05d00cf1e8f999a755151e0451c56ec and elements
[1] 9 4 7 1 2 6 3 8 5
(in some order)
\end{Soutput}
\begin{Sinput}
R> b[b>0.6] <- b[b>0.6] + 3  # add 3 to every element greater than 0.6
R> b
\end{Sinput}
\begin{Soutput}
A disord object with hash 488e1c6f4e2c062379d47b5511730a9785661318 and elements
[1]  5  6 11  4  8  9 12 10  7
(in some order)
\end{Soutput}
\end{Schunk}

Usual semantics follow, provided one is careful to maintain the hash
code:

\begin{Schunk}
\begin{Sinput}
R> d <- disord(1:10)
R> d
\end{Sinput}
\begin{Soutput}
A disord object with hash 65e11d78de79b7f584068ad856749e3748cb837c and elements
 [1]  1  2  3  4  5  6  7  8  9 10
(in some order)
\end{Soutput}
\begin{Sinput}
R> e <- 10 + 3*d - d^2
R> e
\end{Sinput}
\begin{Soutput}
A disord object with hash 65e11d78de79b7f584068ad856749e3748cb837c and elements
 [1]  12  12  10   6   0  -8 -18 -30 -44 -60
(in some order)
\end{Soutput}
\begin{Sinput}
R> e<4
\end{Sinput}
\begin{Soutput}
A disord object with hash 65e11d78de79b7f584068ad856749e3748cb837c and elements
 [1] FALSE FALSE FALSE FALSE  TRUE  TRUE  TRUE  TRUE  TRUE  TRUE
(in some order)
\end{Soutput}
\begin{Sinput}
R> d[e<4] <- e[e<4]
R> d
\end{Sinput}
\begin{Soutput}
A disord object with hash 65e11d78de79b7f584068ad856749e3748cb837c and elements
 [1]   1   2   3   4   0  -8 -18 -30 -44 -60
(in some order)
\end{Soutput}
\end{Schunk}

Above, the replacement command works because {\tt d} and {\tt e}, {\em
and} {\tt e<4} [which is a Boolean {\tt disord} object] all have the
same hash code.

\section{An R session with the {\tt mvp} package}

The {\tt mvp} package implements multivariate polynomials using the
{\tt STL} map class.  Following commands only work as intended here
with {\tt mvp >= 1.0-12}.  Below we see how {\tt disordR} idiom allows
mathematically meaningful operation while suppressing inadmissible
ones:

\begin{Schunk}
\begin{Sinput}
R> set.seed(0)
R> a <- rmvp()
R> b <- rmvp()
R> a
\end{Sinput}
\begin{Soutput}
mvp object algebraically equal to
3 a b^9 e^4 f  +  7 a^2 b^4 d^6 e f^4  +  4 a^4 b^6 c^5 d^11 f^4  +
6 a^6 b^3 c^14 f^2  +  5 a^11 e^6 f^6  +  b^8 e^7 f^12  +  2 b^10 d^10 f^4
\end{Soutput}
\begin{Sinput}
R> b
\end{Sinput}
\begin{Soutput}
mvp object algebraically equal to
5 a c^2 e^8 f^7  +  4 a^2 b^5 c^6 e^3  +  7 a^2 b^7 c^4 d e^2  +
a^4 d^6 e^5 f  +  6 a^6 d^6 f^6  +  3 b^7 c^7 e^5  +  2 b^10 c^3 f^7
\end{Soutput}
\end{Schunk}

Observe that standard multivariate polynomial algebra works:

\begin{Schunk}
\begin{Sinput}
R> a + 2*b
\end{Sinput}
\begin{Soutput}
mvp object algebraically equal to
3 a b^9 e^4 f  +  10 a c^2 e^8 f^7  +  7 a^2 b^4 d^6 e f^4  +
8 a^2 b^5 c^6 e^3  +  14 a^2 b^7 c^4 d e^2  +  4 a^4 b^6 c^5 d^11 f^4  +
2 a^4 d^6 e^5 f  +  6 a^6 b^3 c^14 f^2  +  12 a^6 d^6 f^6  +
5 a^11 e^6 f^6  +  6 b^7 c^7 e^5  +  b^8 e^7 f^12  +  4 b^10 c^3 f^7  +
2 b^10 d^10 f^4
\end{Soutput}
\begin{Sinput}
R> (a+b)*(a-b) == a^2-b^2   # should be TRUE (expression is quite long)
\end{Sinput}
\begin{Soutput}
[1] TRUE
\end{Soutput}
\end{Schunk}

We can extract the coefficients of these polynomials using the
{\tt coeffs()} function:

\begin{Schunk}
\begin{Sinput}
R> coeffs(a)
\end{Sinput}
\begin{Soutput}
A disord object with hash 76b070e3d27bf2e3a548b56a02678d79881de0ce and elements
[1] 3 7 4 6 5 1 2
(in some order)
\end{Soutput}
\begin{Sinput}
R> coeffs(b)
\end{Sinput}
\begin{Soutput}
A disord object with hash 40b9beff42bebe889cb596f78d096c90ef279834 and elements
[1] 5 4 7 1 6 3 2
(in some order)
\end{Soutput}
\end{Schunk}

observe that the coefficients are returned as a {\tt disord}  object.
We may manipulate the coefficients of a polynomial in many ways.  We
may do the following things:

\begin{Schunk}
\begin{Sinput}
R> coeffs(a)[coeffs(a) < 4] <- 0   # set any coefficient of a that is <4 to zero
R> a
\end{Sinput}
\begin{Soutput}
mvp object algebraically equal to
7 a^2 b^4 d^6 e f^4  +  4 a^4 b^6 c^5 d^11 f^4  +  6 a^6 b^3 c^14 f^2  +
5 a^11 e^6 f^6
\end{Soutput}
\begin{Sinput}
R> coeffs(b) <- coeffs(b)%%2       # consider coefficients of b modulo 2
R> b
\end{Sinput}
\begin{Soutput}
mvp object algebraically equal to
a c^2 e^8 f^7  +  a^2 b^7 c^4 d e^2  +  a^4 d^6 e^5 f  +  b^7 c^7 e^5
\end{Soutput}
\end{Schunk}

However, many operations which have reasonable idiom are in fact
meaningless and are implicitly prohibited.  For example:

\begin{Schunk}
\begin{Sinput}
R> x <- rmvp()     # set up new mvp objects x and y
R> y <- rmvp()
\end{Sinput}
\end{Schunk}

Then the following should all produce errors:

\begin{Schunk}
\begin{Sinput}
coeffs(x) + coeffs(y)  # order implementation specific
coeffs(x) <- coeffs(y) # ditto
coeffs(x) <- 1:2       # replacement value not length 1
coeffs(x)[coeffs(x) < 3] <- coeffs(x)[coeffs(y) < 3]
\end{Sinput}
\end{Schunk}

\section{Functions {\tt vars()} and {\tt powers()} return {\tt disord} objects}

The {\tt disord()} function takes a list argument, and this is useful
for working with {\tt mvp} objects:

\begin{Schunk}
\begin{Sinput}
R> (a <- as.mvp("x^2 + 4 - 3*x*y*z"))
\end{Sinput}
\begin{Soutput}
mvp object algebraically equal to
4  -  3 x y z  +  x^2
\end{Soutput}
\begin{Sinput}
R> vars(a)
\end{Sinput}
\begin{Soutput}
A disord object with hash 9395842c6e67dfb0be871e04b3a8964a1c9b9bd5 and elements
[[1]]
character(0)

[[2]]
[1] "x" "y" "z"

[[3]]
[1] "x"

(in some order)
\end{Soutput}
\begin{Sinput}
R> powers(a)
\end{Sinput}
\begin{Soutput}
A disord object with hash 9395842c6e67dfb0be871e04b3a8964a1c9b9bd5 and elements
[[1]]
integer(0)

[[2]]
[1] 1 1 1

[[3]]
[1] 2

(in some order)
\end{Soutput}
\begin{Sinput}
R> coeffs(a)
\end{Sinput}
\begin{Soutput}
A disord object with hash 9395842c6e67dfb0be871e04b3a8964a1c9b9bd5 and elements
[1]  4 -3  1
(in some order)
\end{Soutput}
\end{Schunk}

Note that the hash of all three objects is identical, generated from
the polynomial itself (not just the relevant element of the
three-element list that is an {\tt mvp} object).  This allows us to do
some rather interesting things:

\begin{Schunk}
\begin{Sinput}
R> double <- function(x){2*x}
R> (a <- rmvp())
\end{Sinput}
\begin{Soutput}
mvp object algebraically equal to
a^2 c^10 d^2 f^2  +  7 a^3 d^5 e^14  +  6 a^5 c^7 d^4 e^2  +  4 a^8 c d^5 e^6  +
2 b^2 c^4 d^10 e^5 f  +  5 b^2 c^6 d^2 e^7 f^6  +  3 c^6 d^4 e^2 f^6

\end{Soutput}
\begin{Sinput}
R> pa <- powers(a)
R> va <- vars(a)
R> ca <- coeffs(a)
R> pa[ca<4] <- sapply(pa,double)[ca<4]
R> mvp(va,pa,ca)
\end{Sinput}
\begin{Soutput}
mvp object algebraically equal to
7 a^3 d^5 e^14  +  a^4 c^20 d^4 f^4  +  6 a^5 c^7 d^4 e^2  +
4 a^8 c d^5 e^6  +  5 b^2 c^6 d^2 e^7 f^6  +  2 b^4 c^8 d^20 e^10 f^2  +
3 c^12 d^8 e^4 f^12
\end{Soutput}
\end{Schunk}

Above, {\tt a} was a multivariate polynomial and we doubled the powers
of all variables in terms with coefficients less than 4. Or even:

\begin{Schunk}
\begin{Sinput}
R> a <- as.mvp("3 + 5*a*b - 7*a*b*x^2 + 2*a*b^2*c*d*x*y -6*x*y + 8*a*b*c*d*x")
R> a
\end{Sinput}
\begin{Soutput}
mvp object algebraically equal to
3  +  5 a b  +  8 a b c d x  -  7 a b x^2  +  2 a b^2 c d x y  -  6 x y
\end{Soutput}
\begin{Sinput}
R> pa <- powers(a)
R> va <- vars(a)
R> ca <- coeffs(a)
R> va[sapply(pa,length) > 4] <- sapply(va,toupper)[sapply(pa,length) > 4]
R> mvp(va,pa,ca)
\end{Sinput}
\begin{Soutput}
mvp object algebraically equal to
3  +  8 A B C D X  +  2 A B^2 C D X Y  +  5 a b  -  7 a b x^2  -  6 x y
\end{Soutput}
\end{Schunk}

Above, we took multivariate polynomial {\tt a} and replaced the variable
names in every term with more than four variables with their uppercase
equivalents.

\section{Conclusions}

In the context of {\tt stl} map class, the {\tt disordR} package
allows permissible idiom transparently and without being noticed; but
traps inadmissible constructions with an error.

\bibliographystyle{plain}
\bibliography{disord}

\end{document}